# Understanding longitudinal optical oscillator strengths and mode order

Thomas G. Mayerhöfer[a,c,*], Sonja Höfer[a], Vladimir Ivanovski[b], Jürgen Popp[a,c]

[a] *Leibniz Institute of Photonic Technology (IPHT), Albert-Einstein-Str. 9, D-07745 Jena, Germany*

[b] *Institute of Chemistry, Faculty of Natural Sciences and Mathematics, Sts. Cyril and Methodius University, Arhimedova 5, 1000 Skopje, Macedonia*

[c] *Institute of Physical Chemistry and Abbe Center of Photonics, Friedrich Schiller University, Jena, D-07743, Helmholtzweg 4, Germany*

[*]*Corresponding author. Tel.: +49 (0)3641/948348; Fax: +49 (0)3641/206399; E-mail address: Thomas.Mayerhoefer@ipht-jena.de (T.G. Mayerhöfer)*

**ABSTRACT**

A classical way of describing a dielectric function employs sums of contributions from damped harmonic oscillators. Each term leads to a maximum in the imaginary part of the dielectric function at the transversal optical (TO) resonance frequency of the corresponding oscillator. In contrast, the peak maxima of the negative imaginary part of the inverse dielectric function are attributed to the so-called longitudinal optical (LO) oscillator frequencies. The shapes of the corresponding bands resemble those of the imaginary part of the dielectric function. Therefore, it seems natural to also employ sums of the contributions of damped harmonic oscillators to describe the imaginary part of the inverse dielectric function. In this contribution, we derive the corresponding dispersion relations to investigate and establish the relationship between the transversal and longitudinal optical oscillator strength, which can differ, according to experimental results, by up to three orders of magnitude. So far, these differences are not understood and prevent the longitudinal optical oscillator strengths from proper interpretation. We demonstrate that transversal and longitudinal oscillator strengths should be identical for a single oscillator and that the experimental differences are in this case due to the introduction of a dielectric background in the dispersion formula. For this effect we derive an exact correction. Based on this correction we further derive a modified Kramers-Kronig sum rule for the isotropic case as well as for the components of the inverse dielectric function tensor. For systems with more than one oscillator, our model for the isotropic case can be extended to yield oscillator strengths and LO resonance wavenumber for uncoupled LO modes with or without dielectric background. By comparison with the LO modes obtained from the inverse dielectric function, we find that for oscillators of similar oscillator strength strong coupling results, which exhausts the oscillator strength of the lower LO mode in favor of the higher LO mode, sometimes to the point where the first is hardly traceable. If the oscillator strengths are very different and the TO resonance wavenumber of the weak mode lies between the TO and the LO resonance of the strong mode, an inner and outer mode pair is established. Unlike in monoclinic crystals, for the inner mode pair of the weak mode, the LO resonance wavenumber decreases with increased oscillator strengths. This suggests that from the so-called TO-LO rule, assignments of TO-LO pairs are not reliably possible. In addition, in strongly coupled cases, the LO modes are largely hybridized. Therefore, the meaningfulness of the concept of mode assignment needs to be rethought as well as the recently often employed description of the inverse dielectric function as a sum of independent oscillator terms.

# 1. Introduction

Dispersion theory has a very long tradition that goes back to the 19th century, when Cauchy developed first dispersion formulas which worked well in transparency regions [1]. Astonishingly, the beginning of dispersion analysis, a technique to obtain oscillator parameters from experimental spectra, also goes back to the 19th century, albeit to the end of it, when based on the discovery of anomalous dispersion "modern" dispersion formulas were introduced by Sellmeier [2-3], Helmholtz [4] and Ketteler [5]. In particular, Helmholtz and Ketteler advanced dispersion theory close to the now "classical" form by introducing absorption, even when the derivation was not yet based on Maxwell's equations. This derivation was then introduced by Drude, who also compared the classical to the Maxwell based form and found both equivalent.[6] Planck [7] and Lorentz [8] introduced local field effects in their dispersion theories, which predicted red shifts of the oscillator positions. The classical form of dispersion theory was tested by, among others, Rubens [9]. This test can be seen as the starting point of modern dispersion analysis. Czerny, a former student of Rubens, used it in 1930 to describe the optical properties of NaCl in the infrared spectral range [10]. The term dispersion analysis was coined by Spitzer and Kleinman in the beginning of the 1960ies.[11]

Also in the 1960ies new kinds of dispersion formulas where introduced like the semi-empirical 4-parameter model [12], and a model that takes into account coupling between spectrally neighbored oscillators [13]. About 20 years later anisotropic forms of the classical model were increasingly employed to describe the optical properties of monoclinic and triclinic crystals [14-15] based on theoretical work of Born and Huang [16].

A comparably recent approach, introduced seemingly ad hoc, is the idea that a similar form of the classical dispersion formula can model not only the complex dielectric function itself, but also its inverse. This approach was employed in connection with the analysis of the inverse dielectric function gained by spectroscopic ellipsometry measurements [17]. The probably much more prominent name of the inverse dielectric function is "dielectric loss function". It is in particular of high importance not only for optical, but also for electron energy loss spectroscopy as well as for Raman spectroscopy for vibrational modes that are both, IR and Raman active.[18] In case of the former, the maxima of the negative imaginary part determine the position and the intensity of bands in spectra with *p*-polarized light and high angle of incidence and are related to the well-known Berreman effect, the origin of which is still, more than 60 years of its discovery, under heavy discussion [19-27].

Recently, we used the idea of inverse dielectric function modelling to directly fit reflectance measurements and showed that based on it the problem of determining the oscillator parameter of perpendicular modes (modes that have their transition moment perpendicular to the sample surface) can be elegantly solved and with much less effort than a previously suggested method [28]. Since then, inverse dielectric function modelling has also gained importance in the determination of the longitudinal mode frequency of orthorhombic and monoclinic crystals [29-31].

As its counterpart, the inverse dielectric function model features three parameters per oscillator, which are the oscillator strength, the oscillator position and the damping constant. The latter two are well explained as the longitudinal optical (LO) oscillator frequency which is assumed to be generally blue-shifted relative to the TO frequency due to a net polarization for modes that have their wave vector parallel to this polarization,[32] and the LO-damping constant which can have values different from their transversal optical (TO) counterpart due to mode coupling.[16] What seems to be missing so far is an explanation of the physical meaning of the longitudinal oscillator strengths, the values of which can differ experimentally by more than three orders of magnitude from their TO counterparts [29-31] A phenomenological connection between TO and LO mode amplitudes based on the relation between dielectric tensor function and its inverse has been undertaken very recently, but the physical origin of the deviations remained unexplored and unexplained. [30] While the TO oscillator strength is well-understood and classically a function of the charge, the reduced mass and the number of oscillators per unit volume, the LO oscillator strength is usually not compared to the TO oscillator strengths, let alone, interpreted.

It is therefore the goal of this work to provide such an understanding and to explain the strong deviations between TO and LO oscillator strengths. To that end we first derive the classical damped harmonic oscillator model for the inverse dielectric function and derive an expression for the oscillator strength, which will be shown to be exactly the same as that for the dielectric function itself if only one oscillator is present. Based on Kramers-Kronig sum rules, and further consistency checks, we verify and prove that this finding is indeed the correct result. A large part of the experimental differences can be explained, as we show later, by the consequences of the introduction of a dielectric background into the dispersion formulas. This introduction into the inverse dielectric function model must be performed in a different way compared to the classical model for the dielectric function. Accordingly, the oscillator strength for the TO model must be divided by the square of the dielectric background to resemble its LO counterpart, otherwise the corresponding Kramers-Kronig sum rule is no longer obeyed. Based on this finding, we derive a properly modified Kramers-Kronig sum rule for the inverse dielectric function which takes into account the dielectric background. We further extend our inverse dielectric function model to more than one oscillator and compare the results with those obtained from the inverse dielectric function based on the conventional dielectric function modelling. This comparison will allow to understand coupling effects for LO modes and provides insights to check the so-called TO-LO rule, according to which a TO mode is always followed by its LO mode and which builds the basis for LO mode assignment in recent papers.[31, 33-34] This check reveals that also for crystals with higher symmetry than monoclinic inner and outer TO-LO pairs of oscillators exist and that the LO mode frequency can even become increasingly *smaller* than the TO mode frequency, something which has been demonstrated already 1977 by Gervais.[35] We show that this finding is consistent with experimental results obtained by the semi-empirical 4 parameter model in the older and newer literature. Finally, we discuss the meaningfulness of LO mode assignments and the usefulness of the classical damped harmonic oscillator model for inverse dielectric functions based on our findings.

## 2. Theory

To derive the dispersion relation for the inverse dielectric function, we assume a transverse motion of the atoms in a chain consisting of two alternating kinds of atoms. The motion should be the same in every unit cell consisting of the two different atoms having the reduced mass µ. If we assume damped harmonic oscillators, the (unforced) transversal motion in the unit cell can be described by,

$$\mu\frac{d^2x}{dt^2} + \mu\gamma_{TO}\frac{dx}{dt} + \mu\omega_{TO}^2 x = 0 \:, \tag{1}$$

wherein $\gamma_{TO}$ is the damping constant, $\omega_{TO}$ the eigenfrequency and $x$ the displacement. Eqn. (1) can be derived from Newton's theorem according to which the sum over all forces must vanish at all times. The terms describe in this order acceleration, a damping term proportional to the velocity with an ad-hoc introduced damping constant and a term derived from Hook's law. In the presence of an external field and under negligence of local field effects,[6-8] an additional term $qE$ must be added due to the interaction between the charged atoms having the charge $|q|$ and the electric field $E$:

$$\mu \frac{d^2 x}{dt^2} + \mu \gamma_{TO} \frac{dx}{dt} + \mu \omega_{TO}^2 x = qE , \qquad (2)$$

For longitudinal vibrations (phonon wavevector and polarization have the same direction) in an isotropic crystal, an additional term comes into play, caused by the net polarization introduced by the electric field.[32] This additional term is supposed to increase the resonance frequency from $\omega_{TO}$ to $\omega_{LO}$, so that the following relation results,

$$\mu \frac{d^2 x}{dt^2} + \mu \gamma_{LO} \frac{dx}{dt} + \mu \omega_{LO}^2 x = qE + q P/\varepsilon_0 , \qquad (3)$$

wherein $\gamma_{LO}$ is the LO damping constant, which is for harmonic vibrations equal to $\gamma_{TO}$, but can also be seen as a free parameter for modelling purposes.

Eqn. (3) can be solved following two different paths. The first path is fully analogous to the derivation of the influence of local field effects.[7-8, 36] We take this path first, because we need to determine for the second path the longitudinal dielectric function first and to understand the relation between $\omega_{TO}$ and $\omega_{LO}$, which is essential for the derivation and the discussions provided later on in the manuscript. Accordingly,

$$\mu \frac{d^2 x}{dt^2} + \mu \gamma_{LO} \frac{dx}{dt} + \mu \omega_{LO}^2 x = qE + q P/\varepsilon_0 = qE + q^2 N x/\varepsilon_0 \to$$
$$\mu \frac{d^2 x}{dt^2} + \mu \gamma_{LO'} \frac{dx}{dt} + \mu \left( \omega_{LO}^2 - q^2 N/(\mu \varepsilon_0) \right) x = qE \to \qquad ,$$
$$\mu \frac{d^2 x}{dt^2} + \mu \gamma_{LO'} \frac{dx}{dt} + \mu \omega_{LO'}^2 x = qE$$

(4)

and the polarization will lead to a reduction of $\omega_{LO}$ to $\omega_{LO'}$ just like for the local field of Lorentz.[7-8] To solve eqn. (4), we assume that the displacement $x$ and the electric field $E$ have the same time dependence which is of the form $exp(-i\omega t)$. Thereby we obtain

$$-\mu \omega^2 x - \mu i \gamma_{LO'} \omega + \mu \omega_{LO'}^2 x = qE . \qquad (5)$$

The solution for $x$ is well-known:[4, 6-8]

$$x = \frac{qE}{\mu\left(\omega_{LO'}^2 - \omega^2 - i\gamma_{LO'}\omega\right)}, \tag{6}$$

For non-interacting microscopic dipoles, the macroscopic polarization is equal to $P = Np = NxqE$, where $N$ is the number of dipoles per unit volume. Therefore:

$$P = \frac{Nq^2 E}{\mu\left(\omega_{LO'}^2 - \omega^2 - i\gamma_{LO'}\omega\right)}. \tag{7}$$

We then employ that $P = \varepsilon_0(\varepsilon_r - 1)E$, where $\varepsilon_r$ is the relative dielectric function, divide both sides by $E$ and $\varepsilon_0$ and add unity to both sides to arrive at:

$$\varepsilon_r = 1 + \frac{Nq^2}{\mu\varepsilon_0\left(\omega_{LO'}^2 - \omega^2 - i\gamma_{LO'}\omega\right)}. \tag{8}$$

Setting $S^2 = Nq^2/\mu\varepsilon_0$, with the oscillator strength $S$, we obtain:

$$\varepsilon_r = 1 + \frac{S^2}{\omega_{LO'}^2 - \omega^2 - i\gamma_{LO'}\omega}. \tag{9}$$

How large is the difference between $\omega_{LO}$ and $\omega_{LO'}$? To find an expression for this difference, we use our definition of $\omega_{LO'}$,

$$\omega_{LO'}^2 = \omega_{LO}^2 - S^2, \tag{10}$$

and the expression for the conventional transversal optical dielectric function,

$$\varepsilon_r = 1 + \frac{S^2}{\omega_{TO}^2 - \omega^2 - i\gamma_{TO}\omega}. \tag{11}$$

In the latter, we set $\omega = 0$:

$$\varepsilon_r(0) = 1 + \frac{S^2}{\omega_{TO}^2} \rightarrow \omega_{TO}^2\left(\varepsilon_r(0) - 1\right) = S^2. \tag{12}$$

Furthermore, we use the Lyddane-Sachs-Teller (LST) relation, where we set $\varepsilon_\infty = 1$ and with $\omega_{LO}^2 = \omega_{TO}^2 \varepsilon_r(0)$, we obtain,

$$\omega_{LO}^2 - \omega_{TO}^2 = S^2, \tag{13}$$

and employ this result for $S^2$ in eqn. (10). Accordingly:

$$\omega_{LO'}^2 = \omega_{TO}^2. \tag{14}$$

Obviously, the longitudinal dielectric function is identical to the transversal optical dielectric function, which belongs to the resonance frequency $\omega_{TO}$.

This result is instructive for the interpretation of bands located at the LO-position in the *p*-polarized infrared spectra of isotropic layers and bulk spectra. Our result is in line with the assumption that a minimum or maximum at the LO-position in an infrared spectrum does not automatically mean that an LO-mode has been excited, where we understand by an LO-mode a mode where polarization and wavevector of the phonon have the same direction. In fact, the law of the conservation of momentum tells us that the wave vector of the resulting phonon is not oriented parallel to this polarization and certainly not in cubic materials. Therefore, while the peaks intensities and positions are increasingly determined by the inverse of the dielectric function for higher angles of incidence and *p*-polarized light due to the discontinuity of the perpendicular component of the electric field at an interface, it is still the dielectric function that determines the optical properties,[28] and, in particular, the maximum of its imaginary part, that is characterized by its TO resonance frequency in line with eqn. (14).

As already mentioned, the (longitudinal) dielectric function comes into play, when we consider the second path to go on from eqn. (3). For this path, we employ that $E + P/\varepsilon_0 = \varepsilon_r E$, where $\varepsilon_r$ is the longitudinal relative dielectric function which we just proved to be identical with its transversal counterpart. Hence:

$$\mu \frac{d^2 x}{dt^2} + \mu \gamma_{LO} \frac{dx}{dt} + \mu \omega_{LO}^2 x = q \varepsilon_r E. \tag{15}$$

To solve eqn. (15), we again assume that the displacement *x* and the electric field *E* have the same time dependence and, thereby, arrive at:

$$-\mu \omega^2 x - \mu i \gamma_{LO} \omega + \mu \omega_{LO}^2 x = q \varepsilon_r E. \tag{16}$$

Solving for *x* now results in:

$$x = \frac{q \varepsilon_r E}{\mu \left( \omega_{LO}^2 - \omega^2 - i \gamma_{LO} \omega \right)}. \tag{17}$$

If we again multiply the displacement with the number of oscillators per unit volume *N* and the (effective) charge *q*, we obtain the macroscopic polarization *P*:

$$P = \frac{Nq^2 \varepsilon_r E}{\mu \left( \omega_{LO}^2 - \omega^2 - i\gamma_{LO}\omega \right)} . \tag{18}$$

We then again employ that $P = \varepsilon_0 (\varepsilon_r - 1) E$, divide both sides by $E$ and rearrange the result,

$$\frac{\varepsilon_r - 1}{\varepsilon_r} = 1 - \frac{1}{\varepsilon_r} = \frac{Nq^2}{\mu \varepsilon_0 \left( \omega_{LO}^2 - \omega^2 - i\gamma_{LO}\omega \right)} , \tag{19}$$

from which we obtain by setting $S^2 = Nq^2 / \mu \varepsilon_0$,

$$\varepsilon_r^{-1} = 1 - \frac{S^2}{\omega_{LO}^2 - \omega^2 - i\gamma_{LO}\omega} , \tag{20}$$

and, finally, by converting eqn. (20) to wavenumbers:

$$\varepsilon_r^{-1} = 1 - \frac{S^2}{\tilde{\nu}_{LO}^2 - \tilde{\nu}^2 - i\gamma_{LO}\tilde{\nu}} . \tag{21}$$

From this derivation it is obvious that $S_{LO}^2 = S_{TO}^2 = S^2 = Nq^2 / \mu \varepsilon_0$ and, accordingly, there should be no difference between $S_{LO}$ and $S_{TO}$.

This finding is consistent with the sum rules for the dielectric function and the loss function. These sum rules can be derived on base of the Kramers-Kronig relations (KKR).[37-39] The former sum rule states that the imaginary part of the dielectric function multiplied by the wavenumber is proportional to the squared total oscillator strength:

$$\int_0^\infty \tilde{\nu} \varepsilon_r''(\tilde{\nu}) d\tilde{\nu} = \frac{\pi}{2} S^2 . \tag{22}$$

Since some of the motifs of the derivation are used later on, we provide this derivation shortly in the following. Accordingly, we convert eqn. (11) to the wavenumber form, determine the real part and eliminate in the fraction the factor $\tilde{\nu}_{TO}^2 - \tilde{\nu}^2$:

$$\varepsilon_r' = 1 + \frac{S^2}{\tilde{\nu}_{TO}^2 - \tilde{\nu}^2 + \frac{\tilde{\nu}^2 \gamma^2}{\tilde{\nu}_{TO}^2 - \tilde{\nu}^2}} . \tag{23}$$

Next, we increase the wavenumber $\tilde{\nu}$ to a value (much) higher than the oscillator position, so that $\tilde{\nu}_{TO}^2 \ll \tilde{\nu}^2$. At the same time, $\gamma^2 \ll \tilde{\nu}^2$, so that at this very high wavenumber, the value of the real relative dielectric function is given by,

$$\varepsilon'_r = 1 - \frac{S^2}{\tilde{v}^2} . \tag{24}$$

Alternatively, we can determine $\varepsilon'_r$ from the Kramers-Kronig-Relations,[40]

$$1 - \frac{S^2}{\tilde{v}^2} = 1 + \frac{2}{\pi} \wp \int_0^\infty \frac{\varepsilon''_r(\tilde{v}')\tilde{v}'}{\tilde{v}'^2 - \tilde{v}^2} d\tilde{v}' . \tag{25}$$

wherein $\wp$ indicates the principal value. In the next step we split the integral into two parts by assuming a wavenumber $\tilde{v}_f \ll \tilde{v}$ starting from which $\varepsilon''_r(\tilde{v})$ is practically zero:

$$1 - \frac{S^2}{\tilde{v}^2} = 1 + \frac{2}{\pi} \wp \int_0^{\tilde{v}_f} \frac{\varepsilon''_r(\tilde{v}')\tilde{v}'}{\tilde{v}'^2 - \tilde{v}^2} d\tilde{v}' + \frac{2}{\pi} \wp \int_{\tilde{v}_f}^\infty \frac{\varepsilon''_r(\tilde{v}')\tilde{v}'}{\tilde{v}'^2 - \tilde{v}^2} d\tilde{v}' . \tag{26}$$

As the second integral is effectively zero and because in the first integral the higher limit is much smaller than $\tilde{v}$, we arrive at,

$$\frac{S^2}{\tilde{v}^2} = \frac{2}{\pi} \int_0^{\tilde{v}_f} \frac{\varepsilon''_r(\tilde{v}')\tilde{v}'}{\tilde{v}^2} d\tilde{v}' , \tag{27}$$

from which we can obtain eqn. (22).

Another sum rule has been derived for the loss function:[38]

$$\int_0^\infty \tilde{v} \, \mathrm{Im}(-1/\varepsilon_r(\tilde{v})) d\tilde{v} = \frac{\pi}{2} S^2 . \tag{28}$$

An important consequence of this particular sum rule is that the originally ad-hoc introduced formula for the inverse dielectric function modelling,[17, 29, 41] which would reduce for one harmonic oscillator to,

$$\varepsilon_r^{-1} = 1 + \frac{S^2}{\tilde{v}_{LO}^2 - \tilde{v}^2 - i\gamma_{LO}\tilde{v}} , \tag{29}$$

cannot be correct.

The important difference is the sign before the oscillator term, which must not be positive, since in eqn. (28) the result of the integral would then be negative, but $S^2$ is an always positive quantity (sometimes it is argued that "a mere transformation of the amplitude parameter by a minus sign" can correct the sign before the oscillator term, but the oscillator strength or amplitude comes into play on both sides of eqn. (28), so the "transformation" is cancelled out and the result is still incorrect). Therefore, completely independent of the concrete form of the inverse dielectric function, its imaginary part must always be negative, which can only be accomplished by a negative sign before the oscillator term. This is a further check for the consistency of the derivations presented in this work.

## 3. Results, discussion and further theoretical considerations

All following illustrations are based on calculations carried out with Mathematica. The code comprises in each case only a few lines, which are solely based on the equations and values provided in this manuscript, so that the gentle reader can easily verify all presented results.

First, we will demonstrate the equality of the oscillator strengths used in the harmonic oscillator modelling of the dielectric function as well as of the loss function. To that end, we use a model oscillator with a damping constant $\gamma_{TO} = \gamma_{LO} = 10$ cm$^{-1}$, a TO resonance wavenumber of 1000 cm$^{-1}$ and a LO resonance wavenumber which results from eqn. (13). We numerically evaluate the integrals in eqs. (22) and (28), starting from the dispersion relations eqs. (11) and (21). Eqn. (28) was used for both, the inverse of eqn. (11) (i.e. modelling of the dielectric function $\varepsilon$ itself and then taking the inverse, which we denote in the following as $1/\varepsilon$) and the loss function modelling according to eqn. (21) (denoted as $\varepsilon^{-1}$), between 10 and 5000 cm$^{-1}$. The results are displayed in Figure 1 as well as their percental relative deviations, since these are much smaller than the linewidth of the curves. This is certainly not a surprising result, but a further consistency check and can also be used for comparison with the results in the following two figures.

To summarize, our derivations show what also the sum rules indirectly prove, namely that the oscillator strengths obtained by the modelling of experimental spectra ("dispersion analysis") should be the same, irrespective if the modelling is carried out employing dispersion relations based on eqn. (11) or eqn. (21), at least for one oscillator materials.

If determined by experiment, however, the values of the oscillator strength gained by dielectric function modelling and by inverse dielectric function modelling differ by large amounts, sometimes by three orders of magnitude or more as already mentioned.[29-31] One of the possible reasons is the influence of absorptions in spectral regions higher than the one considered. If the spectral regions, like e.g. the UV-Vis and the infrared, are separated by a transparency range, i.e. a range without absorptions, it is possible to sum up and represent the effect of the absorptions in the spectral region of higher frequency with a constant which is usually termed $\varepsilon_\infty$:

$$\begin{aligned} \varepsilon_r &= \varepsilon_\infty + \frac{S^2}{\tilde{v}_{TO}^2 - \tilde{v}^2 - i\gamma_{TO}\tilde{v}} \quad &\text{(I)} \\ \varepsilon_r^{-1} &= \varepsilon_\infty^{-1} - \frac{S^2}{\tilde{v}_{LO}^2 - \tilde{v}^2 - i\gamma_{LO}\tilde{v}} \quad &\text{(II)} \end{aligned} \qquad (30)$$

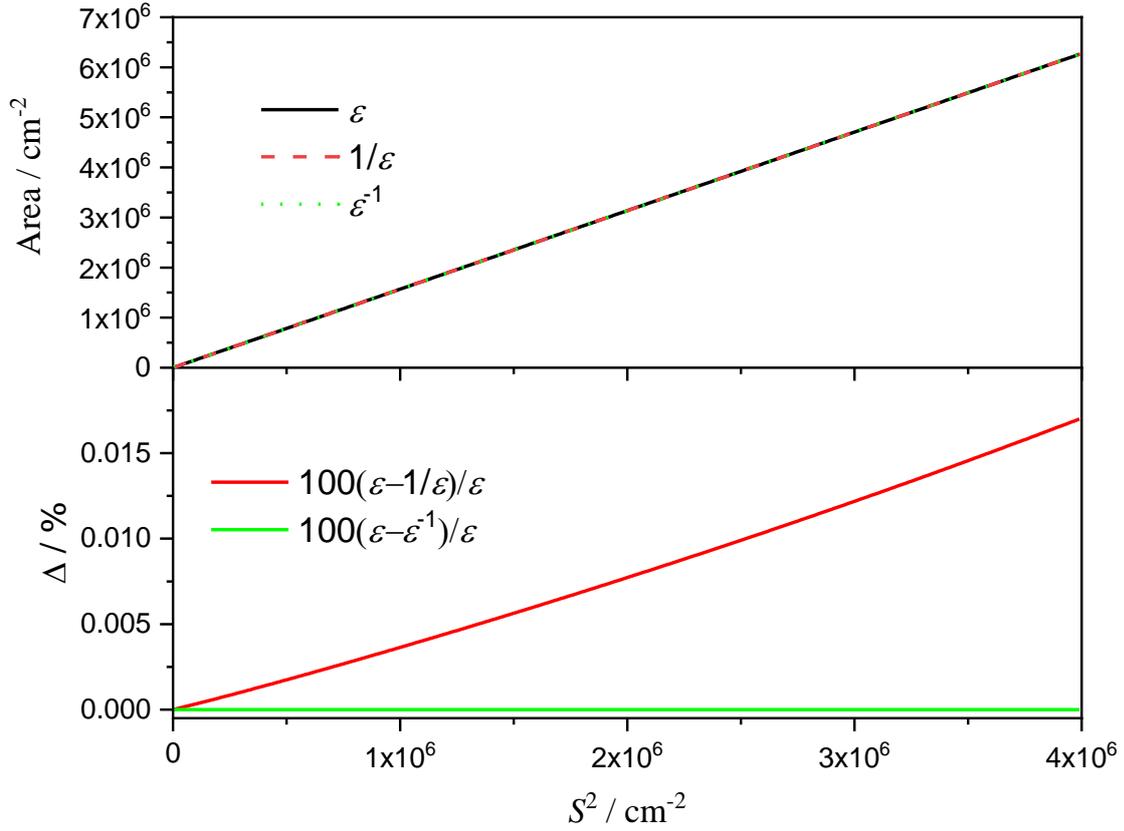

*Figure 1: Upper panel: Integrated model dielectric function ($\varepsilon$), inverse of the model dielectric function ($1/\varepsilon$) and model inverse dielectric function ($\varepsilon^{-1}$) as a function of the oscillator strength $S^2$. Lower panel: Relative differences between $\varepsilon$ and $1/\varepsilon$ (red curve) as well as between $\varepsilon$ and $\varepsilon^{-1}$ (green curve) in percent.*

Eqn. (30), (I) has been tested and verified many times, in contrast to eqn. (30), (II). In fact, eqn. (30), (II) is correct only if $\varepsilon_\infty = 1$. This can easily be verified if the sum rules are applied, which we did for $\varepsilon_\infty = 1.1$ in Figure 2. As is obvious, even an unrealistically small $\varepsilon_\infty$ introduces large, but obviously constant errors.

In fact, as can be verified easily, this constant error is due to a factor $\varepsilon_\infty^2$. If the inverse of eqn. (30) (I) is used, the result of the integration needs to be multiplied by $\varepsilon_\infty^2$ to yield the correct result. In contrast, if eqn. (30) (II) is employed, the result of the integration needs to be divided by $\varepsilon_\infty^2$ to obtain consistent results. This is demonstrated in Figure 3 for which we increased the dielectric background to $\varepsilon_\infty = 3$.

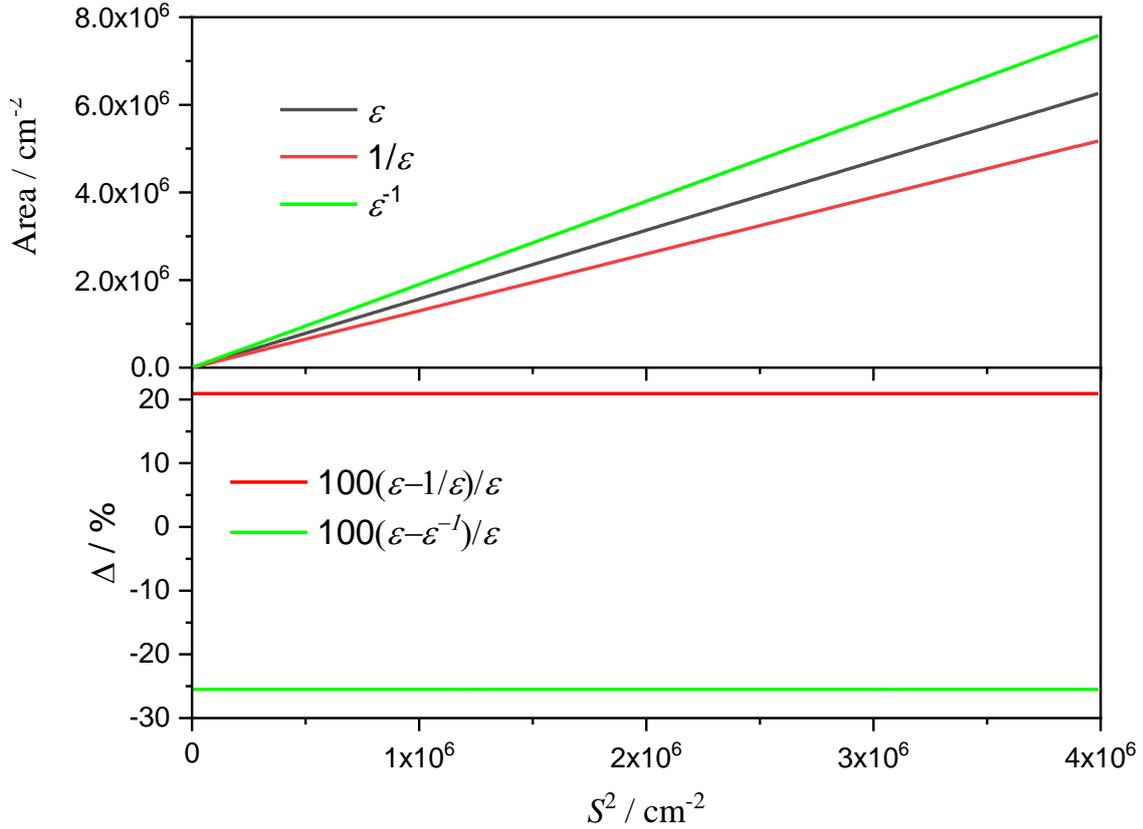

*Figure 2: Upper panel: Integrated model dielectric function ($\varepsilon$), inverse of the model dielectric function ($1/\varepsilon$) and model inverse dielectric function ($\varepsilon^{-1}$) as a function of the oscillator strength $S^2$ if $\varepsilon_\infty = 1.1$. Lower panel: Differences between $\varepsilon$ and $1/\varepsilon$ (red curve) as well as between $\varepsilon$ and $\varepsilon^{-1}$ (green curve) in percent.*

Where does this factor result from? To understand the origin of the factor, we investigate eqn. (30), (II) in the non-absorbing spectral region where the dielectric function and, with it, the inverse dielectric function are real and (nearly) constant. For a wavenumber $\tilde{\nu}_f > \tilde{\nu}_{TO}, \tilde{\nu}_{LO}$, its square is much larger than the squares of the TO and the LO wavenumber: $\tilde{\nu}_f^2 \gg \tilde{\nu}_{TO}^2, \tilde{\nu}_{LO}^2$. Furthermore, $\tilde{\nu}_f^2 \gg \tilde{\nu}_f$, therefore, in equivalence to the derivation of the sum rule, we get:

$$\varepsilon_r = \varepsilon_\infty - \frac{S_{TO}^2}{\tilde{v}_f^2} \quad \text{(I)}$$
$$\varepsilon_r^{-1} = \varepsilon_\infty^{-1} + \frac{S_{LO}^2}{\tilde{v}_f^2} \quad \text{(II)}$$
(31)

If we invert both sides of eqn. (31), (II), we obtain:

$$\varepsilon_r = \frac{1}{\varepsilon_\infty^{-1} - \frac{S_{LO}^2}{-\tilde{v}_f^2}} = \varepsilon_\infty \frac{1}{1 + \frac{\varepsilon_\infty S_{LO}^2}{\tilde{v}_f^2}}. \tag{32}$$

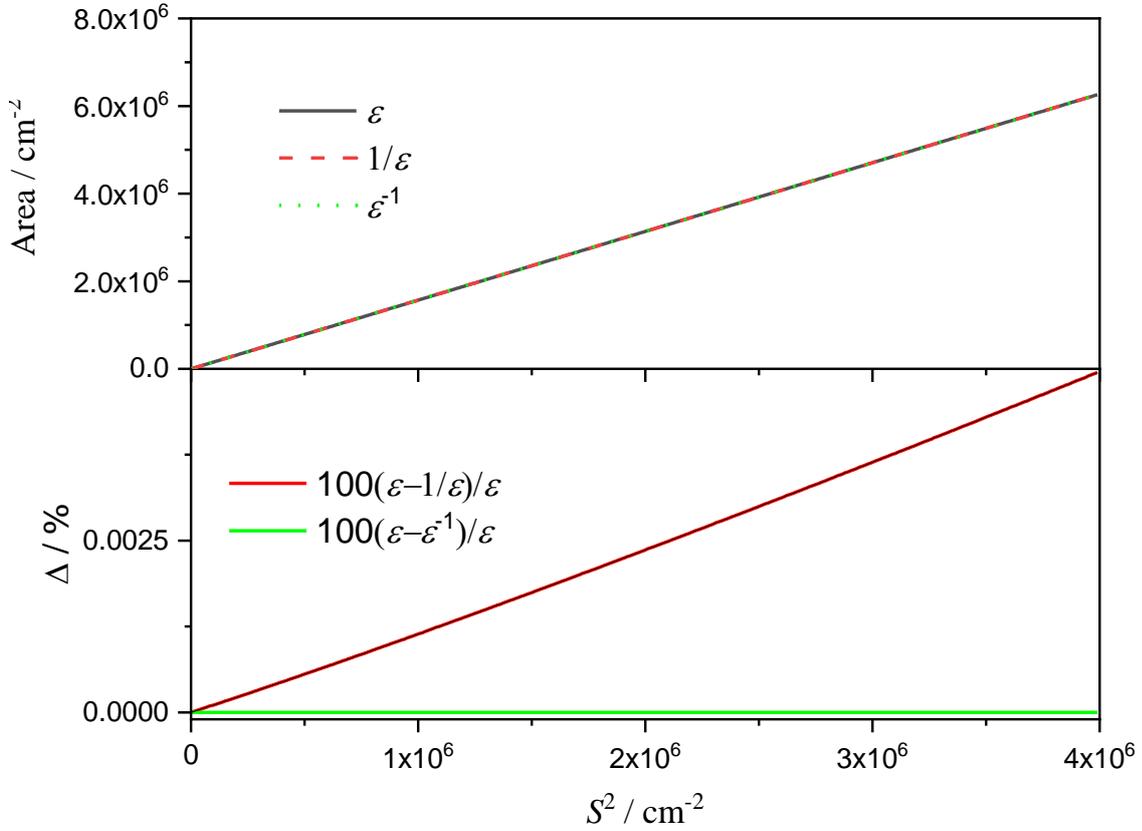

*Figure 3: Upper panel: Integrated model dielectric function ($\varepsilon$), inverse of the model dielectric function ($1/\varepsilon$) and model inverse dielectric function ($\varepsilon^{-1}$) as a function of the oscillator strength $S^2$ if $\varepsilon_\infty = 3$ and the oscillator strength is divided by $\varepsilon_\infty^2$ for the inverse dielectric function modelling, whereas for the inverse of the dielectric function, $S^2$ is multiplied by $\varepsilon_\infty^2$. Lower panel: Differences between $\varepsilon$ and $1/\varepsilon$ (red curve) as well as between $\varepsilon$ and $\varepsilon^{-1}$ (green curve) in percent.*

Using the approximation $1/(1+x) = \sum_{k=0}^{\infty}(-1)^k x^k \approx 1-x$ which is exact for $x \to 0$, where $x = \varepsilon_\infty S_{LO}^2/\tilde{v}_f^2$

and which is valid except for exceptionally strong oscillators, since $S^2 << \tilde{v}_f^2$, the result is:

$$\varepsilon_r = \varepsilon_\infty - \frac{\varepsilon_\infty^2 S_{LO}^2}{\tilde{v}_f^2}. \tag{33}$$

Comparison of eqn. (33) with eqn. (31), (II) reveals that:

$$S^2 = S_{TO}^2 = \varepsilon_\infty^2 S_{LO}^2. \tag{34}$$

For $\varepsilon_\infty = 1$, we find from eqn. (34) that $S_{TO} = S_{LO} = S$, which is a further prove that the two derivations and their results, which we introduced in the beginning of the last section, are consistent.

Figure 3 shows that the approximation introduced by setting $1/(1+x) \approx 1-x$ and the derived relation eqn. (34) is valid at least up to $S_{TO} = 2000 \text{ cm}^{-1}$. Note that this represents an exceptionally strong oscillator, which means that deviations from eqn. (34) are not relevant in practice. In fact, even if it would not make sense from a physical perspective to let $\tilde{v}_f$ increase beyond the transparency region between the infrared and the UV-Vis, mathematically we could choose $\tilde{v}_f$ "as large as" for the derivation of the sum rules. Thereby, $x \to 0$ and relation (34) is proved to be exact.

Furthermore, eqn. (34) also applies to systems of more than one oscillator in equivalence to the Kramers-Kronig sum rules, but only if they are well-separated. Then

$$\varepsilon_r = 1 - \frac{\sum_{j=1}^{N} S_{TO,j}^2}{\tilde{v}_f^2} \quad \text{(I)}$$

$$\varepsilon_r^{-1} = 1 + \frac{\sum_{j=1}^{N} \frac{S_{LO,j}^2}{\varepsilon_{0,j}^2}}{\tilde{v}_f^2} \quad \text{(II)} \tag{35}$$

and

$$S_j^2 = S_{TO,j}^2 = \varepsilon_{0,j}^2 S_{LO,j}^2. \tag{36}$$

In this case, $\varepsilon_{0,j}$ refers to the constant value of the real part of the dielectric function for wavenumbers *lower* than $\tilde{v}_{TO,j}$.

Instead of changing the definition of the oscillator strength according to eqs. (34) and (36), we can alternatively modify the sum rule for the dielectric loss function, eqn. (28), to be valid, independent of the spectral distance of the oscillators, if dispersion relations involving $\varepsilon_\infty$ are employed, to:

$$\int_0^{\tilde{v}_f} \tilde{v} \, \text{Im}(-1/\varepsilon_r(\tilde{v})) d\tilde{v} = \frac{\pi}{2} \frac{S^2}{\varepsilon_\infty^2} \quad S = \sum_{j=1}^{N} S_j. \tag{37}$$

The correctness of eqn. (37) can easily be derived by starting from

$$\varepsilon_r^{-1} = \varepsilon_\infty^{-1} - \sum_{j=1}^{N} \frac{S_j^2/\varepsilon_\infty^2}{\tilde{\nu}_{LO,j}^2 - \tilde{\nu}^2 - i\gamma_{LO,j}\tilde{\nu}}, \tag{38}$$

realizing that for high wavenumbers eqn. (38) is equal to:

$$\varepsilon_r^{-1} = \varepsilon_\infty^{-1} + \frac{1}{\varepsilon_\infty^2 \tilde{\nu}^2} \sum_{j=1}^{N} S_j^2. \tag{39}$$

Therefore

$$\varepsilon_\infty^{-1} + \frac{S^2}{\varepsilon_\infty^2 \tilde{\nu}^2} = \varepsilon_\infty^{-1} + \frac{2}{\pi} \wp \int_0^\infty \frac{\mathrm{Im}(-1/\varepsilon_r)(\tilde{\nu}')\tilde{\nu}'}{\tilde{\nu}'^2 - \tilde{\nu}^2} d\tilde{\nu}', \tag{40}$$

and the result eqn. (37) follows from steps analogously to eqs. (26) and (27).

Correspondingly, for anisotropic materials, the existing Kramers-Kronig sum rule [37] must be modified to,

$$\int_0^{\tilde{\nu}_f} \tilde{\nu} \, \mathrm{Im}(-1/\varepsilon_{r,ij}(\tilde{\nu})) d\tilde{\nu} = \frac{\pi}{2} \frac{S^2}{\varepsilon_{\infty,ij}^2} \delta_{ij} \qquad S = \sum_{j=1}^{N} S_j, \tag{41}$$

wherein $\delta_{ij}$ is the Kronecker symbol. Alternatively to eqn. (41), corresponding corrections of the diagonal elements for the inverse dielectric function tensor could be introduced. For monoclinic and triclinic crystals this correction would, in addition to the determination of the oscillator strengths, also influence that of the orientation of the transition moments due to the anisotropic nature of the correction. Accordingly, the orientation of the longitudinal transition moment determined by inverse dielectric function tensor modelling will be different depending on if or if not the correction is applied.

As already mentioned, the sum rules stay valid even if the oscillators are no longer spectrally distant with a segment of a constant and real dielectric function in between. We use them in the following to investigate and elucidate how the LO oscillator strengths transform in this case. To understand the changes, we take a closer look in the following on two two-oscillator cases. In the first case, the two oscillators have the same TO oscillator strength which is with $S = S_1 = S_2 = 500$ cm$^{-1}$ of medium strength. The first oscillator is stationary at its TO-resonance of 1000 cm$^{-1}$, while we move the second to increasingly higher wavenumbers starting from the same resonance wavenumber as the first. Both oscillators have further in common a damping constant of $\gamma_j = 10$ cm$^{-1}$ and we set $\varepsilon_\infty = 1$.

Not only the peak areas, but, at least roughly, also the peak values follow a simple sum rule if we plot the loss function times the wavenumber over the wavenumber. This becomes obvious in Figure 4 except at narrow spectral distance of the uncoupled LO resonances, where the corresponding peaks increasingly overlap until the oscillator positions coincide and the intensity simply doubles because the two oscillators are no longer distinguishable. It is interesting to observe that the peak value for $\tilde{\nu}_{LO,1}$ is practically zero for resonance wavenumber differences of less than 100 cm$^{-1}$. In this case the coupling between the two oscillators leads to the fact that the first maximum of the inverse dielectric function is practically not observable, in particular also, because it is very close to the TO position. Here, the comparison with the uncoupled LO oscillators accessible through our derivations can greatly help to elucidate the situation. The same is true for the LO-positions where we find strong coupling ("Rabi-splitting" [42]) so that $\tilde{\nu}_{LO,1}$ is red-shifted by about 110 cm$^{-1}$, the same amount by which $\tilde{\nu}_{LO,2}$ is blue-shifted.

Not surprisingly, the larger the spectral distance between both TO-resonances is, the more the LO wavenumbers of the coupled system approach those of the uncoupled oscillators. What is only at the first view surprising, is that even for $\Delta\tilde{\nu}_{TO} = 1000\,\text{cm}^{-1}$, the peak values of the coupled oscillators show about ±20 % deviation from those of the uncoupled case (cf. Fig. 4). The oscillators become spectrally well-separated, which is the situation assumed for eqs. (35) and (36). While the mutual influence on the LO resonance wavenumbers becomes small, the fact remains that the oscillator located at higher wavenumbers changes the dielectric background from unity to $\varepsilon_{0,2} = \tilde{\nu}_{LO,2}^2 / \tilde{\nu}_{TO,2}^2$. Accordingly, the LO oscillator strengths are given by:

$$S_{LO,1} = \left(\frac{\tilde{\nu}_{TO,2}^2}{\tilde{\nu}_{LO,2}^2}\right)^2 S^2; \qquad S_{LO,2} = \left(\frac{\tilde{\nu}_{LO,2}^2}{\tilde{\nu}_{TO,2}^2}\right)^2 S^2. \tag{42}$$

Eqn. (42) can be derived from the already introduced LST-Relation for $\varepsilon_\infty = 1$, $\omega_{LO,j}^2 = \omega_{TO,j}^2 \varepsilon_{0,j}$ together with (36).

We have to keep in mind, that the peak values $P_j$ are only approximately related to the corresponding oscillator strength $S_j$. Nevertheless, as can be seen from Figure 5, eqn. (42) seems to describe the peak values increasingly better the more the coupling vanishes. Accordingly, even in the limiting completely uncoupled case, the oscillator at higher wavenumbers "borrows" oscillator strength from the one at lower wavenumbers.

In the opposite case, where the oscillator strengths are very different, our approach also allows to access and understand some interesting physics which concerns the sequence of the phonon modes. For crystals with higher than monoclinic symmetry, it is generally found that when the wavenumber is increased, a TO resonance is always followed by a LO resonance (TO-LO rule).[43] Accordingly or, better, additionally, it is believed that the sequence is always $\tilde{v}_{TO,1} < \tilde{v}_{LO,1} < \tilde{v}_{TO,2} < \tilde{v}_{LO,2}....$ .[33-34] That this rule is not obeyed for monoclinic crystals, was shown and discussed by Belousov and Pavinich already in 1978,[33] which inspired later works on this subject by some of us [44] and, very recently, in very detailed form, by Schubert et al.[34]

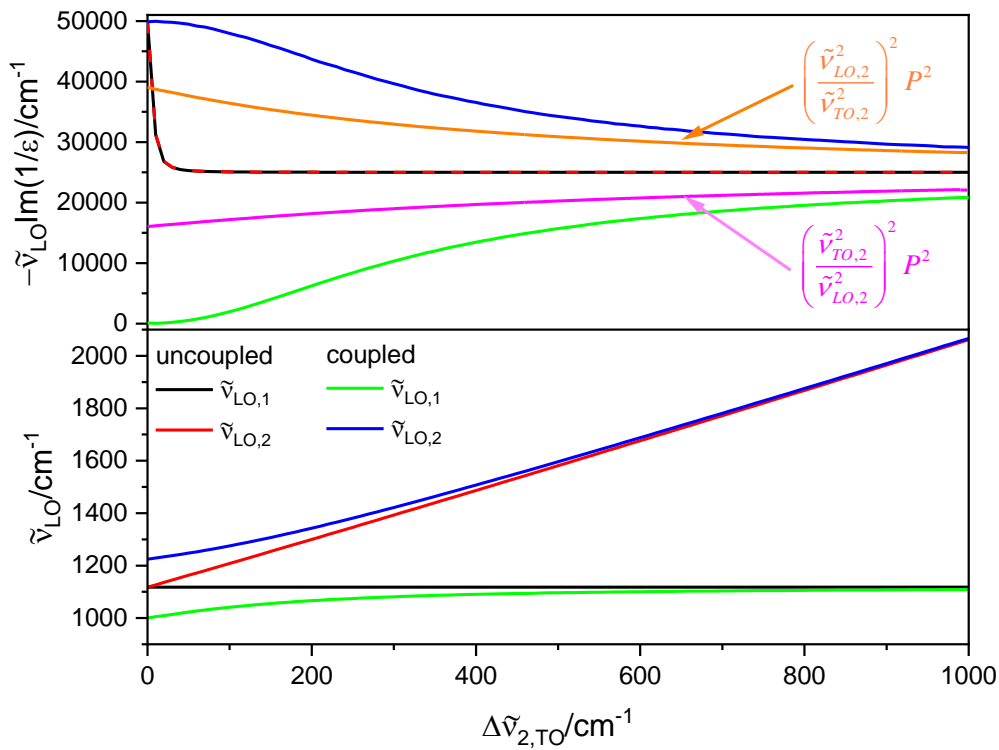

*Figure 4: Two equally strong oscillators with $S_1 = S_2 = 500\,cm^{-1}$. Upper panel: wavenumber times the negative inverse dielectric function at the LO-position for coupled and the uncoupled oscillators. Furthermore, the calculated peak values for uncoupled and spectrally distant oscillators are provided (higher wavenumber mode in orange, lower wavenumber mode in magenta) Lower panel: resonance wavenumbers at the LO-position.*

Equipped with the possibility to understand the LO oscillator strength for uncoupled systems, we found it valuable to reinvestigate the TO-LO rule and, in particular, its interpretation. Such a discussion has already been started by Scott and Porto in 1967 on the example of Raman bands of Quartz [18] and 1977

Gervais continued it.[35] It seems to us that in particular the latter reference did not gain the interest it deserves and stayed mostly unnoticed. We will in the following in particular build upon Gervais' analysis.

To be able to investigate the following example, we also need to understand the change of the LO resonance wavenumber in uncoupled systems with increasing oscillator strength when $\varepsilon_\infty \neq 1$. In these cases, the Lyddane-Sachs-Teller relation is given by $\omega_{LO}^2 \varepsilon_\infty = \omega_{TO}^2 \varepsilon_r(0)$ and with $S^2 = \varepsilon_\infty^2 S_{LO}^2$, we obtain from eqn. (13):

$$\tilde{v}_{LO}^2 = \tilde{v}_{TO}^2 + \frac{S^2}{\varepsilon_\infty}. \tag{43}$$

For the following we assume a strong oscillator located at $\tilde{v}_{TO,1} = 1000$ cm$^{-1}$, having a damping constant of $\gamma_1 = 10$ cm$^{-1}$ and an oscillator strength of $S_1 = 800$ cm$^{-1}$. Assuming $\varepsilon_\infty = 2$, its LO resonance is located at $\tilde{v}_{LO,1} = 1148.9$ cm$^{-1}$. Indeed, if we calculate the dielectric function by putting these parameters into Eqn. (30), (I) and invert it, we find that its negative imaginary part has a maximum at this wavenumber. A second oscillator shall also have a damping constant of $\gamma_2 = 10$ cm$^{-1}$ and its TO resonance at $\tilde{v}_{TO,2} = 1100$ cm$^{-1}$, so that $\tilde{v}_{TO,1} < \tilde{v}_{TO,2} < \tilde{v}_{LO,1}$. If we gradually increase its oscillator strength $S_2$, where is $\tilde{v}_{LO,2}$ supposed to be spectrally located?

For the uncoupled case, eqn. (43) predicts $\tilde{v}_{LO,2}$ to appear at 1100 cm$^{-1}$ and to increase with the square of the oscillator strength divided by $\varepsilon_\infty$ (eqn. (43)). Accordingly, the order would be $\tilde{v}_{TO,1} < \tilde{v}_{TO,2} < \tilde{v}_{LO,2} < \tilde{v}_{LO,1}$, so that inner and outer pairs would also exist for crystals with a symmetry higher than monoclinic. However, the TO-LO rule, stating that each TO mode is followed by a LO mode, would obviously be violated.

In fact, as can be seen from Figure 5, this basic rule is not violated, that is, indeed, one TO-mode is followed by one LO-mode, which is located at higher wavenumber. However, the sequence is in fact $\tilde{v}_{TO,1} < \tilde{v}_{LO,2} < \tilde{v}_{TO,2} < \tilde{v}_{LO,1}$, so that the order of the modes (or better, their mutual assignment) cannot be concluded from the basic TO-LO rule.

Therefore, not only we find an inner and an outer pair, but also that $\tilde{v}_{LO,2} < \tilde{v}_{TO,2}$, because $\tilde{v}_{LO,2}$ is actually *decreasing* with increasing TO oscillator strength while its LO oscillator strengths is increasing in contrast to the previous example. Again, the peak positions and values for the coupled oscillators is gained by calculating the dielectric function, inverting it and determining the values of and at the maxima in dependence of the TO oscillator strength of mode 2. Accordingly, $\tilde{v}_{LO,2}$ follows roughly from the following relation,

$$\tilde{v}_{LO}^2 = \tilde{v}_{TO}^2 - \frac{S^2}{\varepsilon_\infty}, \qquad (44)$$

the values of which are represented by the orange curve in the lower part of Figure 5. At least for $S_{TO,2} < 80$ cm$^{-1}$, the agreement between eqn. (44) and $\tilde{v}_{LO,2}$ is excellent. As can be seen from the peak values of the negative imaginary part multiplied by the LO wavenumber, the oscillators are coupled since the second oscillator "borrows" some intensity from the first, but both oscillators can be seen as to keep mostly their character, so the coupling does not seem to be very strong.

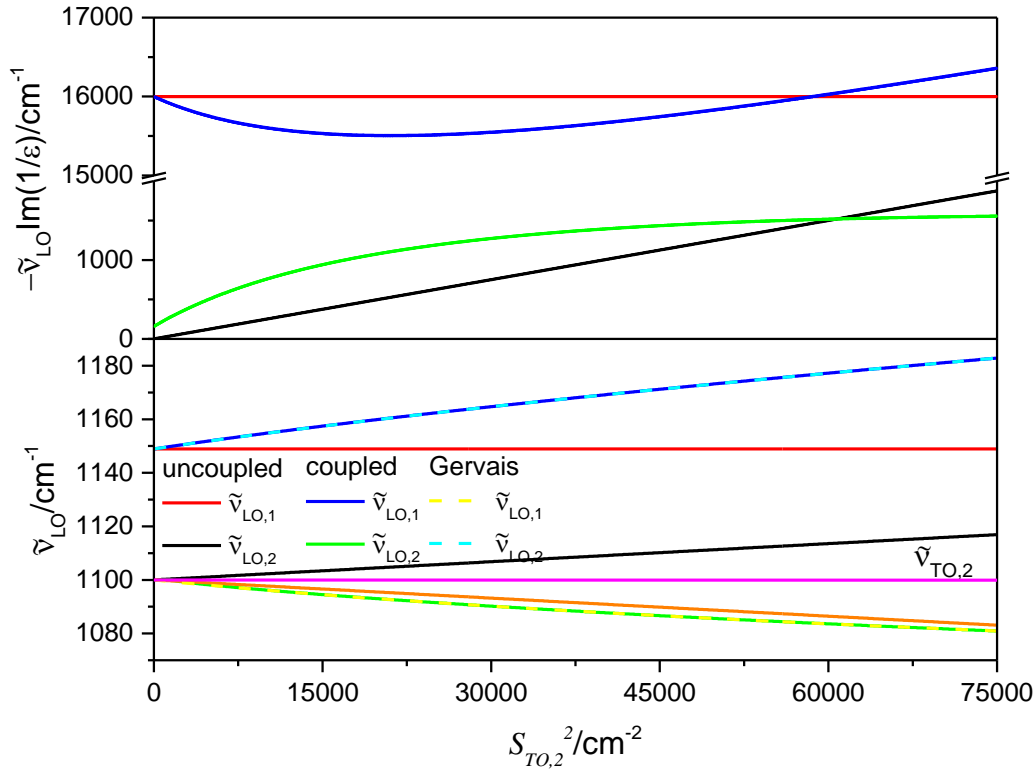

*Figure 5: Strong and weak oscillator case with the sequence $\tilde{v}_{TO,1} < \tilde{v}_{TO,2} < \tilde{v}_{LO,1}$. Upper panel: wavenumber times the negative inverse dielectric function at the LO-position for the two coupled and the two uncoupled oscillators. Lower panel: wavenumber at the LO-position for the uncoupled oscillators (red and black lines), the coupled oscillators gained from $1/\varepsilon$ (blue and green lines) and the Gervais-model (green and cyan lines). The orange line represents the black line mirrored by the magenta line which is the TO-position of the second mode.*

Nevertheless, it is present and a better description of the interdependencies of the LO mode frequencies has been given by Gervais in said reference [35]. Based on the condition $\varepsilon' = 0$, Gervais derived for two coupling oscillators by negligence of damping the following relation:

$$\tilde{N}_{LO\pm}^2 = \frac{1}{2}\left\{\tilde{v}_{LO,1}^2 + \tilde{v}_{LO,2}^2 \pm \left[\left(\tilde{v}_{LO,1}^2 - \tilde{v}_{LO,2}^2\right)^2 + 4\frac{S_1^2 S_2^2}{\varepsilon_\infty^2}\right]^{\frac{1}{2}}\right\}, \tag{45}$$

Here, the $\tilde{v}_{LO,j}$ represent the uncoupled LO-mode resonance positions given by eqn. (43), whereas $\tilde{N}_{LO\pm}$ are the LO-mode resonance positions for the coupled oscillators. Since eqn. (45) neglects damping, it is interesting to find out, how well does it perform if damping is taken into account? To our best knowledge eqn. (45) has never been tested to see how its performance. The result of this test is also displayed in Figure 5. Obviously, eqn. (45) performs excellently and not only for weak modes as the maximal $S \approx 275\,\text{cm}^{-1}$ in Figure 5 and a deviation between damped and undamped case is still not noticeable. Taken this together with the information provided by the comparison of the LO oscillator strengths, we think the mode sequence is proven to be $\tilde{v}_{TO,1} < \tilde{v}_{LO,2} < \tilde{v}_{TO,2} < \tilde{v}_{LO,1}$ where the second LO mode is hybridized (as is the first), but represents largely mode 2.

What is the physical reason behind the decrease of $\tilde{v}_{LO,2}$ with increasing oscillator strength? The polarization induced by the first oscillator is obviously very strong, and since $\tilde{v}_{TO,1} < \tilde{v}_{TO,2} < \tilde{v}_{LO,1}$, the change of the polarization of the first oscillator is not only out of phase but antiparallel to that of the second oscillator, because based on eqs. (3) and (4), this would cause the sign in eqn. (43) to flip and result in eqn. (44).

Examples for the situation $\tilde{v}_{TO,1} < \tilde{v}_{LO,2} < \tilde{v}_{TO,2} < \tilde{v}_{LO,1}$ have been provided amply in literature. In addition to the already mentioned quartz, further examples were found e.g. by Gervais and Pirious in corundum ($E_u$ mode 4) and rutile ($E_u$ mode 2 and 3)[45] and in [46], where the authors reinvestigated quartz and corundum and showed that the case $\tilde{v}_{LO,j} < \tilde{v}_{TO,j}$ also appears for one of the $A_{2u}$ modes of corundum (mode 2), mode 7 of the ordinary ray spectrum of quartz and mode 6 and mode 8 in spinel. A recent example are three very weak modes in orthorhombic NdGaO$_3$, numbered 10-12 in Table 2b of [47]. In all examples, the mode sequence can be safely determined if the semi-empirical 4 parameter model [12, 45, 48] is employed,

$$\varepsilon_i = \varepsilon_{\infty,i} \prod_{j=1}^{N} \frac{\tilde{v}_{LO,j}^2 - \tilde{v}^2 - i\tilde{v}\gamma_{LO,j}}{\tilde{v}_{TO,j}^2 - \tilde{v}^2 - i\tilde{v}\gamma_{TO,j}}, \quad (46)$$

which has been demonstrated in [45-47]. Reinvestigations of corundum,[41] rutile,[49] and NdGaO$_3$ [31] have also been performed recently by spectroscopic ellipsometry. In these papers the authors tacitly or less tacitly reassigned the LO-modes based on the interpretation of the TO-LO rule that each TO mode must be followed by *its* LO mode. Such an interpretation would also contradict the rule of thumb derived from eqs. (43) and (44), and also obeyed by the 4-parameter model, that weak oscillators cannot cause large TO-LO splittings. In this context, also [29] and, specifically, [34] should be reevaluated, in particular also, because the same case of reversed order of the LO and TO resonance wavenumber can also occur for *p*-polarized spectra due to mode interaction between modes of different symmetry [50-51].

Furthermore, while those have not been investigated with the semi-empirical 4-parameter model so far, it is obvious that many of the 2 atomic materials that crystallize in the cubic crystal system like LiF, NaCl, MgO etc.[10, 52] will also show this type of mode sequence with the fundamental building up the outer and the first harmonic establishing the inner mode pair.

In all cases, comparably small dips in broad reflectance plateaus are concerned, with the exception of mode 6 in spinel which can, due to its mode strength, be only considered as large dip. For all these cases, however, $\tilde{v}_{TO,j} < \tilde{v}_{TO,j+1} < \tilde{v}_{LO,j}$ applies and obviously leads to the sequence $\tilde{v}_{TO,j} < \tilde{v}_{LO,j+1} < \tilde{v}_{TO,j+1} < \tilde{v}_{LO,j}$.

A further conclusion of cases where $\tilde{v}_{LO} < \tilde{v}_{TO}$ is that

$$S^2 = \varepsilon_\infty \left| \tilde{v}_{LO}^2 - \tilde{v}_{TO}^2 \right| \quad (47)$$

must replace eqn. (43), otherwise oscillators with $\tilde{v}_{LO} < \tilde{v}_{TO}$ would make negative contributions and violate the sum rules, which obviously they don't do. In the generalized Lyddane-Sachs-Teller relation,[53]

$$\frac{\varepsilon_0}{\varepsilon_\infty} = \prod_{j=1}^{N} \frac{\tilde{v}_{LO,j}^2}{\tilde{v}_{TO,j}^2}, \quad (48)$$

however, assignment errors certainly do not play a role, since the position of factors can be exchanged without changing the result.

One last thing that we investigated in connection with the case $\tilde{v}_{LO} < \tilde{v}_{TO}$ was if the rule derived in [41],

$$\sum_{j=1}^{N}\left(\gamma_{LO,j} - \gamma_{TO,j}\right) > 0, \qquad (49)$$

which shall assure that $\text{Im}\{\varepsilon\} > 0$ if the semi-empirical 4-parameter model is employed, is somehow affected by the LO-TO inversion. To that goal, we used the same oscillator parameter as used in the previous model (the LO-mode positions were determined by eqn. (45): $\tilde{v}_{LO,1} = 1182.85 \, \text{cm}^{-1}$ and $\tilde{v}_{LO,2} = 1080.91 \, \text{cm}^{-1}$) and set $\gamma_{LO,1} = \gamma_{TO,1}$ while varying $\gamma_{LO,2}$. In a second run, we set $\gamma_{LO,2} = \gamma_{TO,2}$ and varied $\gamma_{LO,1}$. The minimal values of $\text{Im}\{\varepsilon\}$ in the range between 800 and 1500 cm$^{-1}$ are displayed in Figure 6.

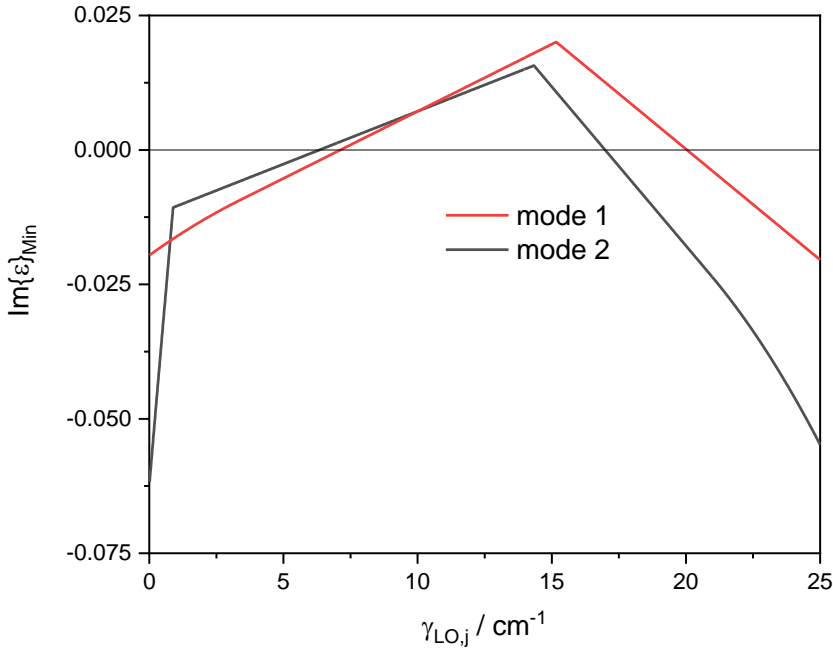

*Figure 6: Minimum value of the imaginary part of the dielectric function when the semi-empirical 4-parameter model is employed and one strong and one weak oscillator is assumed with the TO-mode positions $\tilde{v}_{TO,1} = 1000$ cm$^{-1}$ and $\tilde{v}_{TO,2} = 1100$ cm$^{-1}$ and the LO-mode positions $\tilde{v}_{LO,1} = 1182.85$ cm$^{-1}$ and $\tilde{v}_{LO,2} = 1080.91$ cm$^{-1}$. The TO-damping parameters were both set to $\gamma_{TO,1} = \gamma_{TO,2} = 10$ cm$^{-1}$. One of the LO-damping parameters was also fixed at 10 cm$^{-1}$ and the other was varied.*

According to (49), $\text{Im}\{\varepsilon\} > 0$ should be fulfilled when the other LO-damping parameter is larger than 10, a condition that is obviously not valid in the case shown in Figure 6, where the region were $\text{Im}\{\varepsilon\} > 0$, is given by 7.13 cm$^{-1} < \gamma_{LO,1} < 20.03$ cm$^{-1}$ and 6.34 cm$^{-1} < \gamma_{LO,2} < 16.99$ cm$^{-1}$. For mode 2, it could be expected that $\gamma_{LO,2} \approx \gamma_{TO,2}$, but the first mode is strong, so that larger differences between $\gamma_{LO,1}$ and $\gamma_{TO,1}$ can be expected. While in our experience for spectral fitting with the semiempirical 4-paramter model introducing a penalty function in line with (49) works quite well, even when sometimes weaker modes take on larger values of $\gamma_{LO}$ to compensate lower values of $\gamma_{LO}$ for stronger modes [28], care obviously must be taken when modes with $\tilde{\nu}_{LO} < \tilde{\nu}_{TO}$, since then condition eqn. (49) is not valid. Generally, it might be put into question if it really is always meaningful to assign a LO mode to a particular TO mode. As emphasized several times, TO modes in reflectance or spectroscopic ellipsometry spectra can be described usually very well by the classical oscillator model, since for TO modes usually no strong coupling exists. For LO modes on the other hand, commonly strong coupling can be found, at least in inorganic materials, and the modes are mostly of hybrid nature as already pointed out by Gervais.[35] It is therefore a legitimate subject for discussion, if a description as sums of the contribution of individual terms of the oscillators is meaningful at all. Instead it might be more useful, also in light of the findings above, to characterize the inverse dielectric function by the semi-empirical 4-parameter model as this was suggested in [29]:

$$\varepsilon_i^{-1} = \varepsilon_{\infty,i}^{-1} \prod_{j=1}^{N} \frac{\tilde{\nu}_{TO,j}^2 - \tilde{\nu}^2 - i\tilde{\nu}\gamma_{TO,j}}{\tilde{\nu}_{LO,j}^2 - \tilde{\nu}^2 - i\tilde{\nu}\gamma_{LO,j}}, \qquad (50)$$

For $\tilde{\nu}_{TO,j}$ and $\gamma_{TO,j}$, the values gained from the analysis of the dielectric function could be used, and LO-oscillator strengths obtained from eqn. (47). Those values could be compared with those obtained by the uncoupled LO oscillator model introduced in this work. From the comparison, the hybrid character of the LO-modes could be derived and erroneous or meaningless assignments could be avoided.

## 5. Summary and Conclusion

We presented a derivation of the inverse dielectric function / loss function dispersion based on the conventional derivation of the classical damped harmonic oscillator model. Thereby we gained the originally ad-hoc introduced inverse dielectric function model with the important difference that we correct a sign error present in this ad-hoc model. Furthermore, we find that in the infrared spectral region the dielectric background caused by absorptions in higher spectral regions is not only represented as a constant which is added to the term contributed by the oscillator, but also alters the values that are obtained for the oscillator strength by a multiplicative factor for which we derived that it is given by the squared dielectric background. Based on these results we suggested a modified sum rule applicable for spectral regions where a dielectric background is useful. Starting from the formulas derived for one oscillator, we investigate the changes introduced by a second oscillator. For modes of similar strength, we found by comparison with the uncoupled case, a very strong coupling of the LO modes which spectrally extends over several hundreds of wavenumbers. If a comparably weak oscillator is introduced and has its TO wavenumber in between the TO and LO wavenumber of a strong mode, the comparison of coupled and uncoupled model shows that an inner and an outer pair is formed and that the LO wavenumber of the weaker oscillator actually decreases with increasing oscillator strengths. This proves that from the TO-LO rule a phonon mode assignment cannot be reliably obtained. Generally, mode coupling of the LO modes is strong, so it is questionable if the inverse dielectric function can be meaningfully described by a summation of contributions – the inverted semi-empirical form might be better suited for this task.

Overall, we think that with the derivations and conclusions presented in this work provide an important step towards a full understanding of the nature and the benefits of inverse dielectric function modelling, in particular also with regard to the Berreman effect and of the connection between TO and LO modes in general.

# References

1. Kayser, H., *Handbuch der Spektroskopie, Vol. 4*. Verlag von S. Hirzel: 1908.
2. Sellmeier, W., *Annalen der Physik* **1872,** *223* (11), 386-403.
3. Sellmeier, W., *Annalen der Physik* **1872,** *223* (12), 525-554.
4. Helmholtz, H., *Annalen der Physik* **1875,** *230* (4), 582-596.
5. Ketteler, E., *Annalen der Physik* **1887,** *266* (2), 299-316.
6. Drude, P., *Annalen der Physik* **1893,** *284* (3), 536-545.
7. Planck, M., *Sitzungsberichte der Königlich Preussischen Akademie der Wissenschaften* **1902,** *I*, 470-494.
8. Lorentz, H. A., *Koninkl. Ned. Akad. Wetenschap. Proc.* **1906,** *8*, 591-611.
9. Rubens, H., *Annalen der Physik* **1892,** *281* (2), 238-261.
10. Czerny, M., *Z. Physik* **1930,** *65* (9-10), 600-631.
11. Spitzer, W.; Kleinman, D., *Physical Review* **1961,** *121* (5), 1324-1335.
12. Berreman, D. W.; Unterwal.Fc, *Physical Review* **1968,** *174* (3), 791-&.
13. Barker, A.; Hopfield, J., *Physical Review* **1964,** *135* (6A), A1732-A1737.
14. Pavinich, V. F.; Belousov, M. V., *Optics and spectroscopy* **1978,** *45* (6), 881-3.
15. Emslie, A. G.; Aronson, J. R., *J. Opt. Soc. Am.* **1983,** *73* (7), 916-919.
16. Born, M.; Huang, K., *Dynamical Theory of Crystal Lattices*. Clarendon Press: 1954.
17. Humlíček, J., *Philosophical Magazine Part B* **1994,** *70* (3), 699-710.
18. Scott, J. F.; Porto, S. P. S., *Physical Review* **1967,** *161* (3), 903-&.
19. Yasuhisa, K.; Takamaro, K., *Japanese Journal of Applied Physics* **2009,** *48* (12R), 121406.
20. Giovanna, S.; Jeong-Seok, N.; Gregory, N. P., *Journal of Physics: Condensed Matter* **2010,** *22* (15), 155401.
21. Shaganov, I. I.; Perova, T. S.; Melnikov, V. A.; Dyakov, S. A.; Berwick, K., *The Journal of Physical Chemistry C* **2010,** *114* (39), 16071-16081.
22. Raman, R.; Mishra, P.; Kapoor, A. K.; Muralidharan, R., *Journal of Applied Physics* **2011,** *110* (5), 053519.
23. Ishitani, Y., *Journal of Applied Physics* **2012,** *112* (6), -.
24. Neubrech, F.; Pucci, A., *Ieee Journal of Selected Topics in Quantum Electronics* **2013,** *19* (3).
25. Chalopin, Y.; Hayoun, M.; Volz, S.; Dammak, H., *Applied Physics Letters* **2014,** *104* (1), 011905.
26. Huck, C.; Vogt, J.; Neuman, T.; Nagao, T.; Hillenbrand, R.; Aizpurua, J.; Pucci, A.; Neubrech, F., *Optics express* **2016,** *24* (22), 25528-25539.
27. Berte, R.; Gubbin, C. R.; Wheeler, V. D.; Giles, A. J.; Giannini, V.; Maier, S. A.; De Liberato, S.; Caldwell, J. D., *ACS Photonics* **2018,** *5* (7), 2807-2815.
28. Mayerhöfer, T. G.; Ivanovski, V.; Popp, J., *Spectrochimica acta. Part A, Molecular and biomolecular spectroscopy* **2016,** *168*, 212-7.
29. Mock, A.; Korlacki, R.; Knight, S.; Schubert, M., *Physical Review B* **2018,** *97* (16), 165203.
30. Schubert, M.; Mock, A.; Korlacki, R.; Knight, S.; Galazka, Z.; Wagner, G.; Wheeler, V.; Tadjer, M.; Goto, K.; Darakchieva, V., *Applied Physics Letters* **2019,** *114* (10), 102102.
31. Mock, A.; Korlacki, R.; Knight, S.; Stokey, M.; Fritz, A.; Darakchieva, V.; Schubert, M., *Physical Review B* **2019,** *99* (18), 184302.
32. Kittel, C., *Introduction to Solid State Physics*. Wiley: 2004.
33. Belousov, M. V.; Pavinich, V. F., *Optika I Spektroskopiya* **1978,** *45* (5), 920-926.
34. Schubert, M.; Mock, A.; Korlacki, R.; Darakchieva, V., *Physical Review B* **2019,** *99* (4).

35. Gervais, F., *Optics Communications* **1977,** *22* (1), 116-118.
36. Born, M., *Optik: Ein Lehrbuch der elektromagnetischen Lichttheorie*. Julius Springer: 1933.
37. Altarelli, M.; Dexter, D. L.; Nussenzveig, H. M.; Smith, D. Y., *Physical Review B* **1972,** *6* (12), 4502-4509.
38. Wooten, F., *Optical Properties of Solids*. Elsevier Science: 2013.
39. Tanner, D. B., *Optical Effects in Solids*. Cambridge University Press: 2019.
40. Mayerhöfer, T. G.; Popp, J., *Spectrochimica Acta Part A: Molecular and Biomolecular Spectroscopy* **2019,** *213*, 391-396.
41. Schubert, M.; Tiwald, T. E.; Herzinger, C. M., *Physical Review B* **2000,** *61* (12), 8187-8201.
42. Limonov, M. F.; Rybin, M. V.; Poddubny, A. N.; Kivshar, Y. S., *Nature Photonics* **2017,** *11*, 543.
43. Kurosawa, T., *Journal of the Physical Society of Japan* **1961,** *16* (7), 1298-1308.
44. Ivanovski, V.; Mayerhöfer, T. G.; Popp, J., *Vibrational Spectroscopy* **2007,** *44* (2), 369-374.
45. Gervais, F.; Piriou, B., *Journal of Physics C: Solid State Physics* **1974,** *7* (13), 2374.
46. Zeidler, S.; Posch, T.; Mutschke, H., *A&A* **2013,** *553*, A81.
47. Höfer, S.; Uecker, R.; Kwasniewski, A.; Popp, J.; Mayerhöfer, T. G., *Vibrational Spectroscopy* **2015,** *78*, 17-22.
48. Gervais, F.; Piriou, B., *Physical Review B* **1974,** *10* (4), 1642-1654.
49. Schöche, S.; Hofmann, T.; Korlacki, R.; Tiwald, T. E.; Schubert, M., *Journal of Applied Physics* **2013,** *113* (16), -.
50. Ivanovski, V.; Petruševski, V. M., *J. Mol. Struct.* **2003,** *650* (1–3), 165-173.
51. Ivanovski, V.; Petruševski, V. M., *Spectrochimica Acta Part A: Molecular and Biomolecular Spectroscopy* **2004,** *60* (7), 1601-1607.
52. Jasperse, J. R.; Kahan, A.; Plendl, J. N.; Mitra, S. S., *Physical Review* **1966,** *146* (2), 526-542.
53. Cochran, W.; Cowley, R. A., *Journal of Physics and Chemistry of Solids* **1962,** *23* (5), 447-450.